\documentclass{article}

\usepackage[preprint,nonatbib]{neurips_2020}

\usepackage[utf8]{inputenc} 
\usepackage[T1]{fontenc}    
\usepackage{hyperref}       
\usepackage{url}            
\usepackage{booktabs}       
\usepackage{amsfonts}       
\usepackage{pdfpages}
\usepackage{nicefrac}       
\usepackage{amsmath}
\usepackage{xcolor}
\usepackage{graphicx}
\usepackage{microtype}  
\usepackage{enumitem}
\newcommand{\etal}{\textit{et al}.~}
\usepackage{subcaption}

\DeclareUnicodeCharacter{FFFC}{~}
\DeclareUnicodeCharacter{308}{~}

\newcommand\todo[1]{\textcolor{green}{#1}}

\title{Improved Conditional Flow Models for \\
Molecule to Image Synthesis}

\author{%
  Karren Yang\thanks{Massachusetts Institute of Technology, Cambridge, MA, USA;}\;\,\footnotemark[3]
  \And
  Samuel Goldman\footnotemark[1]
  \And 
  Wengong Jin\footnotemark[1]
  \And 
  Alex Lu\thanks{University of Toronto, Toronto ON, Canada;}
  \AND
  Regina Barzilay\footnotemark[1]
  \And
  Tommi Jaakkola\footnotemark[1]
  \And
  Caroline Uhler\footnotemark[1]\;\,\thanks{To whom correspondence should be addressed: karren@mit.edu, cuhler@mit.edu}
}

\begin{document}

\maketitle

\begin{abstract}
In this paper, we aim to synthesize cell microscopy images under different molecular interventions, motivated by practical applications to drug development. Building on the recent success of graph neural networks for learning molecular embeddings and flow-based models for image generation, we propose Mol2Image: a flow-based generative model for molecule to cell image synthesis. To generate cell features at different resolutions and scale to high-resolution images, we develop a novel multi-scale flow architecture based on a Haar wavelet image pyramid. To maximize the mutual information between the generated images and the molecular interventions, we devise a training strategy based on contrastive learning. To evaluate our model, we propose a new set of metrics for biological image generation that are robust, interpretable, and relevant to practitioners. We show quantitatively that our method learns a meaningful embedding of the molecular intervention, which is translated into an image representation reflecting the biological effects of the intervention.
\end{abstract}

\section{Introduction}
High-content cell microscopy assays are gaining traction in recent years as the rich morphological data from the images proves to be more informative for drug discovery than conventional targeted screens \cite{caicedo2016applications, eggert2013and, swinney2011were}. Motivated by these developments, we aim to build, to our knowledge, the first generative model to synthesize cell microscopy images under different molecular interventions, translating molecular information into a high-content and interpretable image representation of the intervention. Such a system has numerous practical applications in drug development -- for example, it could enable practitioners to virtually screen compounds based on their predicted morphological effects on cells, allowing more efficient exploration of the vast chemical space and reducing the resources required to perform extensive experiments \cite{reddy2007virtual,shoichet2004virtual,walters1998virtual}. In contrast to conventional models that predict specific chemical properties, a molecule-to-image synthesis model has the potential to produce a panoptic view of the morphological effects of a drug that captures a broad spectrum of properties such as mechanisms of action \cite{ljosa2013comparison,loo2009approach,perlman2004multidimensional} and gene targets \cite{breinig2015chemical}. 

To build our molecule-to-image synthesis model (Mol2Image), we integrate state-of-the-art graph neural networks for learning molecular representations with flow-based generative models. Flow-based models are a relatively recent class of generative models that learn the data distribution by directly inferring the latent distribution and maximizing the log-likelihood of the data \cite{dinh2014nice, dinh2016density, kingma2018glow}. Compared to other classes of deep generative models such as variational autoencoders (VAEs) \cite{kingma2013auto} and generative adversarial networks (GANs) \cite{goodfellow2014generative}, flow-based models do not rely on approximate posterior inference or adversarial training and are less prone to training instability and mode collapse, making them advantageous for biological applications \cite{sun2019dual}. %
%
Nevertheless, molecule-to-image synthesis is a challenging task that highlights key, unsolved problems in flow-based generation.
Current flow architectures cannot scale to full-resolution cell images (e.g., $512 \times 512$) and are unable to separately generate image features at multiple spatial resolutions, which is important to disentangle coarse features (e.g., cell distribution) from fine features (e.g., subcellular localization of proteins). While separate generation of image features at different resolutions has been demonstrated using GANs~\cite{denton2015deep}, this is still an open problem for flow-based models. Furthermore, existing formulations of flow models do not effectively leverage conditioning information when the relationship between the image and the conditioning information is complex and/or subtle as is the case with molecular interventions. This results in generated samples that do not reflect the conditioning information.

\textbf{Contributions.} In this work, we develop (to our knowledge) the first molecule-to-image synthesis model for generating high-resolution cell images conditioned on molecular interventions. Specifically,
\vspace{-0.5cm}
\begin{itemize}[leftmargin=*]
\itemsep0em 
    \item We develop a new architecture and approach for flow-based generation based on a Haar wavelet image pyramid, which generates image features at different spatial resolutions separately and enables scaling to high-resolution images.
    \item We propose a new training strategy based on contrastive learning to maximize the mutual information between the latent variables of the flow model and the embedding of the molecular graph, ensuring that generated images reflect the molecular intervention.
    \item We establish a set of evaluation metrics specific to biological image generation that are robust, interpretable, and relevant to biological practitioners.
    \item We demonstrate that our method outperforms the baselines by a large margin, indicating potential for application to virtual screening.
\end{itemize}

Although we focus on molecule-to-image synthesis in this work, our generative approach can potentially extend to other applications, e.g., text-to-image synthesis \cite{reed2016generative}.

\begin{figure}[t]

\begin{subfigure}{.6\textwidth}
  \centering
  \includegraphics[width=1\linewidth]{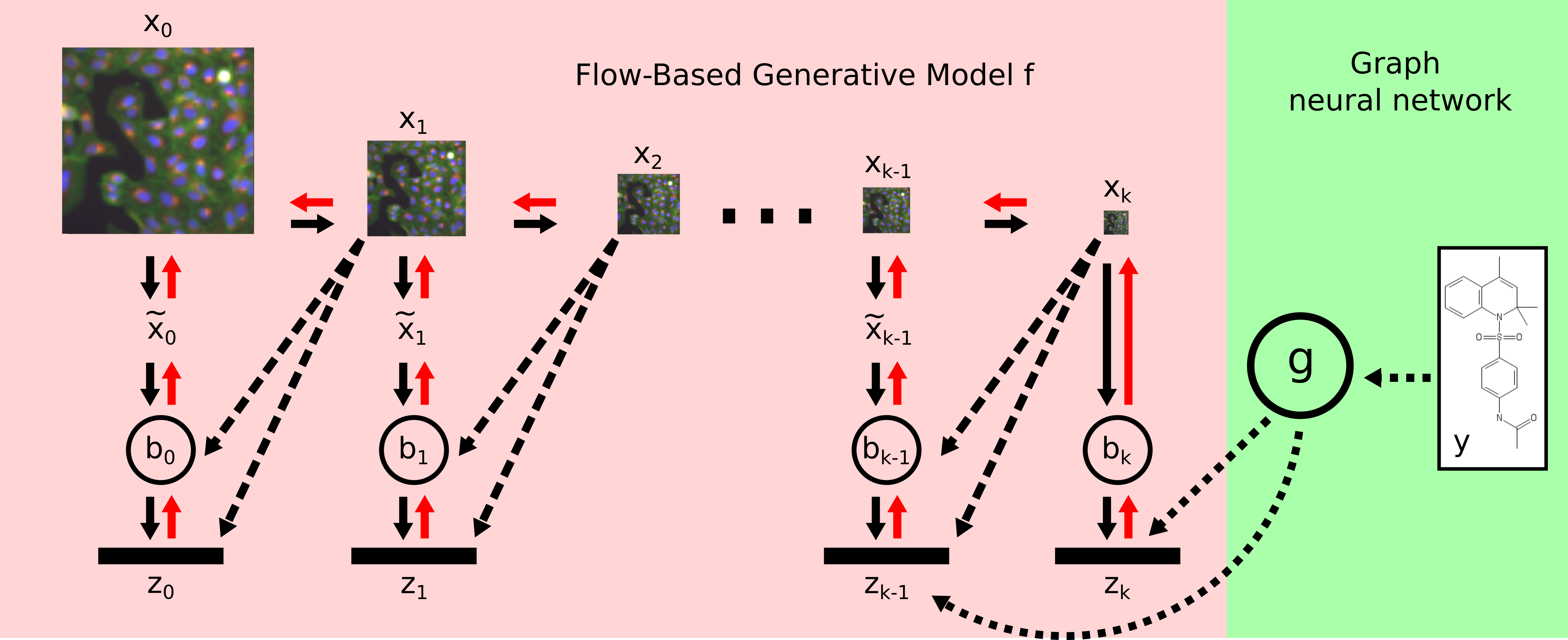}  
  \caption{Model Architecture}
  \label{fig:architecture}
\end{subfigure}
\quad
\begin{subfigure}{.4\textwidth}
  \centering
  \includegraphics[width=1\linewidth]{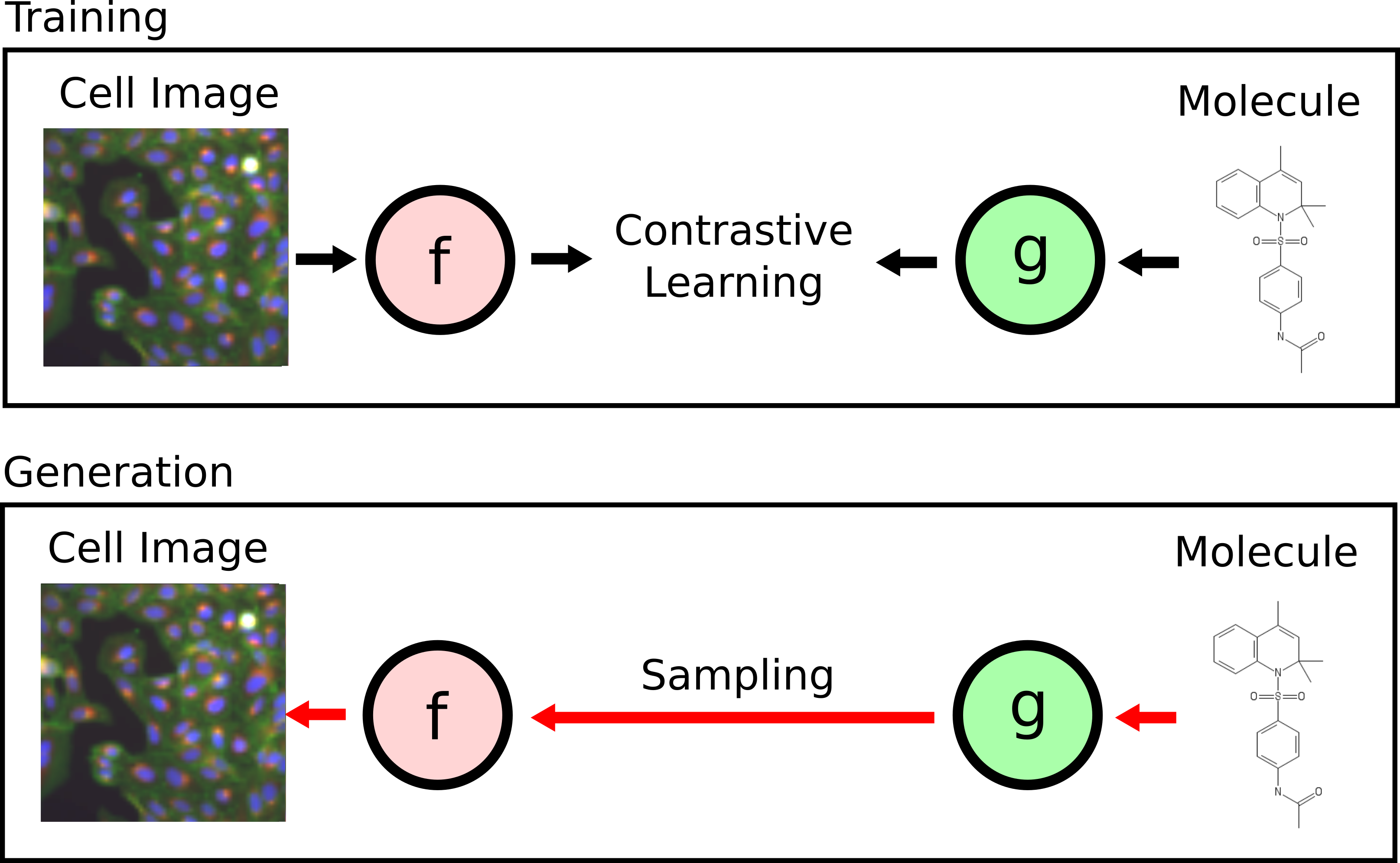}  
  \caption{Training Strategy}
  \label{fig:contrastive-loss}
\end{subfigure}
\caption{(a) (Red box) Our flow-based model architecture based on a Haar wavelet image pyramid. Information flow follows the black arrows during training/inference and the red arrows during generation. The dashed lines represent conditioning and are used in both training and generation. (Green box) Molecular information is processed and input to the network via a graph neural network $g$. (b) Our training strategy for effective molecule-to-image synthesis. See text for details.}
\label{fig:method}
\end{figure}


\section{Related Work}

\textbf{Biological Image Generation.}
Osokin \etal use GAN architectures to generate cellular images of budding yeast to infer missing fluorescence channels (stained proteins) in a dataset where only two channels can be observed at a time \cite{osokin2017gans}. Separately, Goldsborough \etal qualitatively evaluate the use of different GAN variants in generating three-channel images of human breast cancer cell lines \cite{goldsborough2017cytogan}. While these works consider the task of generating single cell images, neither considers the generation of cells conditioned on complex inputs nor the generation of multi-cell images, which is useful in observing cell-to-cell interactions \cite{okagaki2010cryptococcal} and variability \cite{mattiazzi2020systematic}. A separate, similar line of investigation in histopathology and medical imagery has used GAN models to refine and generate synthetic datasets for training downstream classifiers but does not address the difficulty of conditional image generation necessary to capture drug interventions \cite{hou2017unsupervised, mahmood2018unsupervised, yi2019generative}. While both high throughput image-based drug screens \cite{caicedo2018weakly} and molecular structures \cite{yang2019analyzing} have been used to generate representations of small molecules, little work has focused on 
learning representations of these modalities jointly.

\textbf{Graph Neural Networks for Molecules.}
A neural network formulation on graphs was first proposed by Gori \etal\cite{gori2005graph}; Scarselli \etal\cite{scarselli2009graph} and later extended to various graph neural network (GNN) architectures~\cite{li2015gated,dai2016discriminative,niepert2016learning,kipf2016semi,hamilton2017inductive,lei2017deriving,velickovic2017graph,xu2018powerful}. In the context of molecule property prediction, Duvenaud \etal\cite{duvenaud2015convolutional} and Kearns \etal\cite{kearnes2016molecular} first applied GNNs to learn neural fingerprints for molecules. Gilmer \etal\cite{gilmer2017neural} further enhanced GNN performance by using set2set readout functions and adding virtual nodes into molecular graphs. Yang \etal\cite{yang2019analyzing} provided extensive benchmarking of various GNN architectures and demonstrated the advantage of GNNs over traditional Morgan fingerprints~\cite{rogers2010extended} as well as domain-specific features. While these works mainly focused on predicting numerical chemical properties, we here focus on using GNNs to learn rich molecular representations for molecule-to-image synthesis.

\textbf{Flow-Based Generative Models.} A flow-based generative model (e.g., Glow) is a sequence of invertible networks that transforms the input distribution to a simple latent distribution such as a spherical Gaussian \cite{dinh2014nice, dinh2016density, ho2019flow++, kingma2018glow, louizos2017multiplicative, rezende2015variational}. Conditional variants of Glow have recently been proposed for image segmentation \cite{lu2019structured, winkler2019learning}, modality transfer \cite{kondo2019flow, sun2019dual}, image super-resolution \cite{winkler2019learning}, and image colorization \cite{ardizzone2019guided}. These applications are variants of image-to-image translation tasks and leverage the spatial correspondence between the conditioning information and the generated image. Other conditional models perform generation given an image class \cite{kingma2018glow} or a binary attribute vector \cite{liu2019conditional}. Since the condition is categorical, these models apply auxiliary classifiers in the latent space to ensure that the model learns the correspondence between the condition and the image. Unlike these works, we generate images from molecular graphs; here spatial correspondence is not present and the conditioning information cannot be learned using a classifier. Therefore we must leverage other techniques to ensure correspondence between the generated images and the conditioning information.

In addition to conditioning on molecular structure, our flow model architecture is based on an image pyramid, which conditions the generation of fine features at a particular spatial resolution on a coarse image from another level of the pyramid. Flow-based generation of images conditioned on other images has been explored in various previous works \cite{ardizzone2019guided, kondo2019flow, lu2019structured, sun2019dual, winkler2019learning}, but different from these works, our flow-based model leverages conditioning to break generation into successive steps and refine features at different scales. Our approach is inspired by methods such as Laplacian Pyramid GANs \cite{denton2015deep} that break GAN generation into successive steps. A key design choice here is our use of a Haar wavelet image pyramid instead of a Laplacian pyramid, which avoids introducing redundant variables into the model and is an important consideration for flow-based models. Ardizzone \etal~\cite{ardizzone2019guided} use the Haar wavelet transform to improve training stability, but they do not consider the framework of an image pyramid for separately generating features at different spatial resolutions. 

\section{Method}
Our approach is to develop a flow-based generative model for synthesizing cell images conditioned on the molecular embeddings of a graph neural network. We first provide an overview of graph neural networks (Section \ref{sec:gnn}) and generative flows (Section \ref{sec:flow}). In Section \ref{sec:architecture}, we describe our novel multi-scale flow architecture that generates images in a coarse-to-fine process based on the framework of an image pyramid. The architecture separates generation of image features at different spatial resolutions and scales to high-resolution cell images.
In Section \ref{sec:contrastive-loss}, we describe a novel training strategy using contrastive learning for effective molecule-to-image synthesis.

\subsection{Preliminaries: Graph Neural Networks} \label{sec:gnn}
A molecule $y$ can be represented as a labeled graph $\mathcal{G}_y$ whose nodes are the atoms in the molecule and edges are the bonds between the atoms. Each node $v$ has a feature vector $\mathbf{f}_v$ including its atom type, valence, and other atomic properties. Each edge $(u,v)$ is also associated with a feature vector $\mathbf{f}_{uv}$ indicating its bond type. 
A graph neural network (GNN) $g$ learns to embed a graph $\mathcal{G}_y$ into a continuous vector $g(y)$. In this paper, we adopt the GNN architecture from \cite{dai2016discriminative,yang2019analyzing}, which associates hidden states $\mathbf{h}_v$ with each node $v$ and updates these states by passing messages $\mathbf{m}_{\vec{uv}}$ over edges $(u,v)$. Each message $\mathbf{m}_{\vec{uv}}^{(0)}$ is initialized at zero. At time step $t$, the messages are updated as follows:
\begin{equation}
    \mathbf{m}_{\vec{uv}}^{(t+1)} = \mathrm{MLP}\big(\mathbf{f}_u, \mathbf{f}_{uv}, \sum_{w \in N(u),w\neq v}\nolimits \mathbf{m}_{\vec{wu}}^{(t)}\big) \quad \forall (u,v) \in \mathcal{G}_y
\end{equation}
where $N(u)$ is the set of neighbor nodes of $u$ and $\mathrm{MLP}$ stands for a 
multilayer perceptron. After $T$ message passing steps, we compute hidden states $\mathbf{h}_v$ as well as the final representation $g(y)$ as
\begin{equation}
    \mathbf{h}_u = \mathrm{MLP}\big(\mathbf{f}_u, \sum_{v \in N(u)}\nolimits \mathbf{m}_{\vec{uv}}^{(t)}\big) \qquad g(y) = \mathrm{MLP}(\sum_{u \in \mathcal{G}_y} \nolimits \mathbf{h}_u)
\end{equation}

\subsection{Preliminaries: Generative Flows} \label{sec:flow}
A generative flow $f$ consists of a sequence of invertible functions $f = f_1 \circ \cdots \circ f_L$ that transform an input variable $\mathbf{x}$ to a Gaussian latent variable $\mathbf{z}$. The generative process is defined as:
\begin{equation} 
\mathbf{z} \sim \mathcal{N}(\mu, \Sigma), \quad \mathbf{h}_L =\mathbf{z}, ~\mathbf{h}_{L-1} = f^{-1}_L(\mathbf{h}_L), ~\cdots, ~\mathbf{h}_0 = f^{-1}_1(\mathbf{h}_1), \quad ~\mathbf{x} = \mathbf{h}_0,
\end{equation}
where $\{h_i\}_{i \in 1 \cdots L}$ are the intermediate variables that arise from applying the inverse of individual flow functions $\{f_i\}_{i \in 1 \cdots L}$. By the change-of-variables formula, the log-likelihood of sampling $\mathbf{x}$ is,
\begin{equation}\label{eq:logl}
\log p(\mathbf{x}) = \log p_\mathcal{N}(\mathbf{z}; \mu, \Sigma) + \sum_{i=1}^L \log\big|\text{det}\frac{d\mathbf{h}_i}{d\mathbf{h}_{i-1}}\big|,
\end{equation}
where $p_\mathcal{N}$ is the Gaussian probability density function. In this paper, we adopt the flow functions from the Glow model \cite{kingma2018glow}, in which each flow consists of actnorm, $1 \times 1$ convolution, and coupling layers (see \cite{kingma2018glow} for details). The Jacobian matrices of these transformations are triangular and hence have log-determinants that are easy to compute. As a result, the log-likelihood of the data is tractable and can be efficiently optimized with respect to the parameters of the flow functions.


\subsection{Proposed Multi-Scale Flow Architecture} \label{sec:architecture}
Existing multi-scale architectures for generative flows \cite{dinh2016density, kingma2018glow} do not separately generate features for different spatial resolutions 
and cannot scale to full-resolution cell images. In the following, we propose a novel multi-scale architecture that generates cell images in a coarse-to-fine fashion and enables scaling to high-resolution images. Our architecture integrates flow units into the framework of an image pyramid generated by recursive 2D Haar wavelet transforms. 

\textbf{Haar Wavelets.} Wavelets are functions that can be used to decompose an image into coarse and fine components. The Haar wavelet transform generates the coarse component in a way that is equivalent to nearest neighbor downsampling. The coarse component is obtained by convolving the image with an averaging matrix followed by sub-sampling by a factor of 2, and the fine components are obtained by convolving the image with three different matrices followed by sub-sampling by a factor of 2:
\begin{equation} \label{eq:haar-matrices}
    M_\textrm{average} = 
\begin{bmatrix}
1 & 1 \\
1 & 1 
\end{bmatrix},
M_\textrm{diff1} = 
\begin{bmatrix}
1 & -1 \\
1 & -1 
\end{bmatrix},
M_\textrm{diff2} = 
\begin{bmatrix}
1 & 1 \\
-1 & -1 
\end{bmatrix},
M_\textrm{diff3} = 
\begin{bmatrix}
1 & -1 \\
-1 & 1 
\end{bmatrix}.
\end{equation}

To generate an image pyramid that captures features at different spatial resolutions, we recursively apply Haar wavelet transforms to the coarse image. Specifically, let $[x_0, x_1, \cdots, x_k]$ be a pyramid of downsampled images, where $x_i$ represents the image $x_0$ after $i$ applications of the coarse operation. We apply the fine operation to each downsampled image except the last, resulting in the image pyramid $[\tilde{x}_0, \tilde{x}_1, \cdots, \tilde{x}_{k-1}, x_k]$. The image at each spatial resolution can be reconstructed recursively,
$$ x_i = I([U(x_{i+1}), \tilde{x}_i]),$$
where $U$ represents spatial upsampling, the brackets indicate concatenation, and $I$ represents the inverse of the linear operation corresponding to the 2D Haar wavelet transform; see Equation (\ref{eq:haar-matrices}).

\textbf{Haar Pyramid Generative Flow.}
Our flow architecture $f$ consists of multiple blocks $b_0, \cdots, b_k$, each responsible for generating the fine features for a different level of the Haar image pyramid conditioned on a coarse image from the next image in the pyramid; see Figure \ref{fig:architecture}, red box. Note that each block $b_i$ consists of multiple invertible flow units, i.e., $b_i = f_1^{(i)} \circ \cdots \circ f_L^{(i)}$ and can be treated independently as a generative flow from Section \ref{sec:flow}. The generative process is defined as follows. First we generate the final downsampled image of the pyramid,
\begin{equation}
     z_k \sim \mathcal{N}(\mu_k, \Sigma_k), \quad x_k = b_k^{-1}(z_k),
\end{equation}
by sampling a latent vector that corresponds to the coarsest features and passing it through the first block. Then we recursively sample latent vectors corresponding to finer spatial features and generate the other images in the Haar image pyramid as follows:
\begin{equation*}
    z_i \sim \mathcal{N}(\mu_i(x_{i+1}), \Sigma_i(x_{i+1})), \quad
    \tilde{x}_i = b_i^{-1}(z_i, x_{i+1}), \quad 
    x_i = I([U(x_{i+1}), \tilde{x}_i]), \quad 0 \leq i < k,
\end{equation*}
where $x = x_0$ is the final full-resolution image. 
To perform conditioning on the coarse image $x_{i+1}$, we provide it as an additional input to both the prior distribution of $z_i$ and to the individual flow units in $b_i$. 
Computation of the log-likelihood within the image pyramid framework is straightforward, since the Haar wavelet transform is an invertible linear transformation with a block-diagonal Jacobian matrix that adds a constant factor to the log-determinant in Equation (\ref{eq:logl}). 

\textbf{Conditioning on a Molecular Graph.}
To condition the generation of features by block $b_i$ on a molecular intervention $y$, we condition the distribution of latent variables $z_i$ on the output of a graph neural network. Specifically, we let $\mu_i, \Sigma_i$ take $g(y)$ as input, where $g$ is a graph neural network described in Section \ref{sec:gnn}; see Figure \ref{fig:contrastive-loss}, green box.

\subsection{Training Strategy: Maximizing Mutual Information using Contrastive Learning}  \label{sec:contrastive-loss}

The challenge of training a conditional flow model using log-likelihood is that it may not sufficiently leverage the shared information between the input image and the molecular intervention. Intuitively, the flow model can achieve a high log-likelihood by converting the input image distribution to a Gaussian distribution without using the condition. This is especially true for molecule-to-image synthesis because the effect of the molecular intervention on the cells is subtle in the image space. 

To ensure that the conditional flow model extracts useful information from the molecular graph for generation, we propose a training strategy based on \emph{contrastive learning}. As shown in Figure \ref{fig:contrastive-loss}, during training, we use contrastive learning to maximize the mutual information between the latent variables from the flow model $f$ and the molecular embedding from the graph neural network $g$. During generation, information flow is reversed through the flow model to generate an image that is tightly coupled to the conditioning molecular information.

The objective of contrastive learning is to learn embeddings of $\mathbf{x}$ and $\mathbf{y}$ that maximize their mutual information. Specifically, these embeddings should distinguish ``matched" samples from the joint distribution $p_{xy}$ from ``mismatched" samples from the product of the marginals $p_x p_y$. To obtain these embeddings, we train a critic function $h$ to assign high values to matched samples and low values to mismatched samples by minimizing the following contrastive loss:
\begin{equation}\label{eq:contrastive_loss}
    \mathcal{L}_\textrm{contrastive} = 
    - \mathbb{E}_{(x_1,y_1) \sim p_{xy}, y_2 \cdots y_N \sim p_y} \left[ \log \frac{h(x_1, y_1)}{\sum_{i=1}^N h(x_1, y_i)} \right].
\end{equation}

In practice, we compute $h(x,y)$ by taking the cosine similarity of $f(x)$ and $g(y)$, where $f$ is the flow model and $g$ is the graph neural network that embeds the molecular structure graph $y$:
\begin{equation}
    h(x, y) = \text{exp} \left( \frac{f(x)\cdot g(y)}{\tau ||f(x)|| \cdot ||g(y)||} \right),
\end{equation}
where $\tau > 0$ is a temperature hyperparameter. Minimizing the contrastive loss in Equation (\ref{eq:contrastive_loss}) is equivalent to maximizing a lower bound on the mutual information between $f(x)$ and $g(y)$ and has been used in previous work for representation learning \cite{oord2018representation}. Our key insight is in leveraging contrastive learning in a conditional flow model to maximize the mutual information between the latent image variables $f(x)$ and the molecular embedding $g(y)$, such that reversing information flow through $f$ generates images that share a high degree of information with the molecular graph $y$. 

\begin{figure}
    \centering
    \includegraphics[scale=0.08]{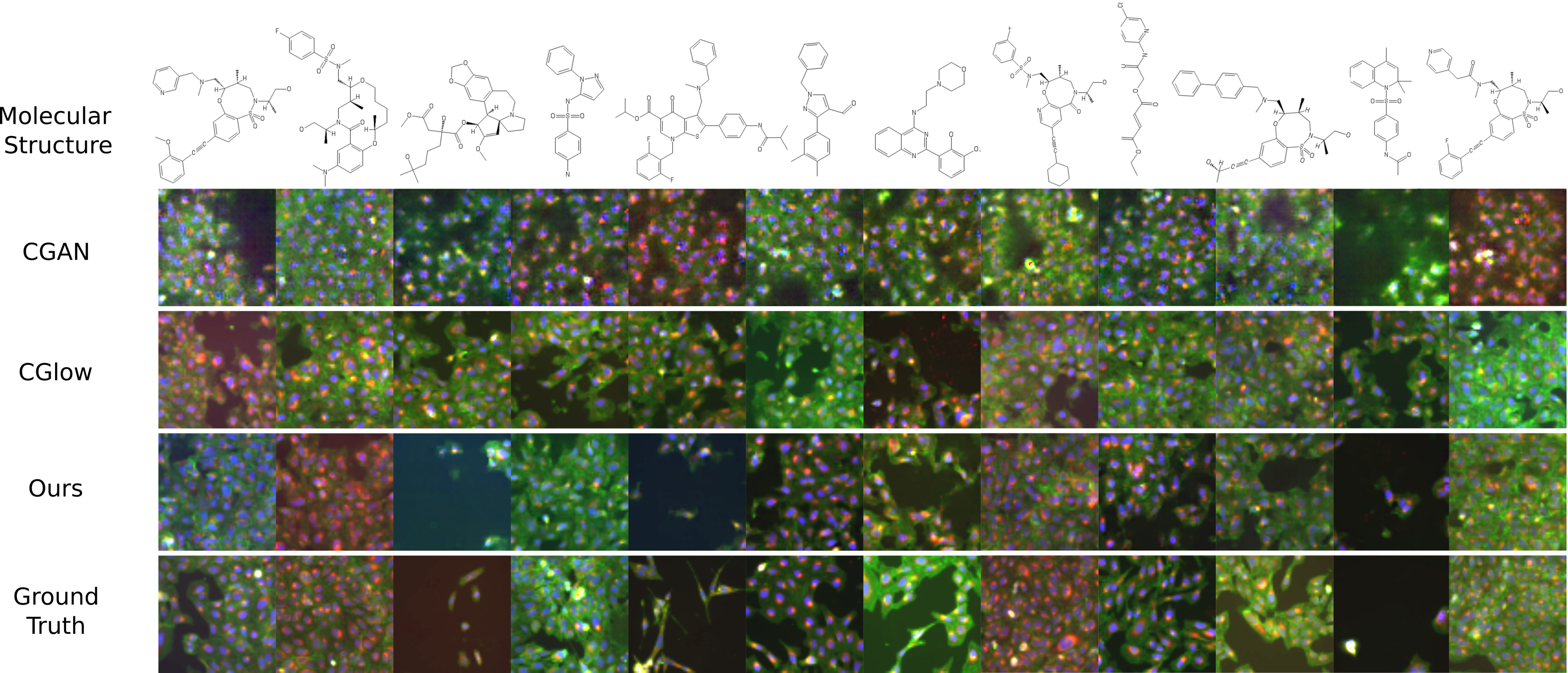}
    \caption{Examples of cell images generated by our method vs the baselines.}
    \label{fig:generated-examples}
\end{figure}

\section{Experiments}
\textbf{Dataset.} We perform our experiments on the Cell Painting dataset introduced by Bray \etal \cite{bray2017dataset,bray2016cell} and preprocessed by Hofmarcher \etal \cite{hofmarcher2019accurate}. The dataset consists of 284K cell images collected from 10.5K molecular interventions. We divide the dataset into a training set of 219K images corresponding to 8.5K molecular interventions, and hold out the remaining of the data for evaluation. The held-out data consists of images corresponding to each of the 8.5K molecules in the training set as well as images corresponding to 2K molecules that are not in the training set.

\textbf{Implementation and Training Details.} Our model for the molecule-to-image generation task consists of six flow modules that construct different levels of the Haar wavelet image pyramid, generating images from resolution of $16 \times 16$ to $512 \times 512$. The lowest resolution module consists of 64 flow units, and each of the other modules consists of 32 flow units. Each of the modules is trained to maximize the log-likelihood of the data (Equation \ref{eq:logl}). Additionally, the three flow modules that process low-resolution images (up to $64 \times 64$ resolution) are also trained to maximize the mutual information between the latent variables and the molecular features using contrastive learning with a weight of $0.1$ and $\tau=0.07$. We train each flow module for approximately 50K iterations using Adam \cite{kingma2014adam} with initial learning rate of $10^{-4}$, during which the highest resolution block sees over 1M images and the lowest resolution block sees over 10M images.

\textbf{Robust Evaluation Metrics for Biological Image Generation.} For a molecule-to-image synthesis model to be useful to practitioners, it needs to generate image features that are meaningful from a biological standpoint. It has been shown that machine learning methods can discriminate between microscopy images using features that are irrelevant to the target content \cite{shamir2011assessing, lu2019cells}. Therefore, in addition to more conventional vision metrics, we propose a new set of evaluation metrics based on CellProfiler cell morphology features \cite{mcquin2018cellprofiler} that are more robust, interpretable, and relevant to practitioners \cite{rohban2017systematic}. We specifically consider the following morphological features:
\vspace{-0.1cm}
\begin{itemize}[leftmargin=*]
\itemsep0em 
\item \textbf{Coverage.} The total area of the regions covered by segmented cells.
\item \textbf{Cell/Nuclei Count.} The total number of nuclei/cells found in the image.
\item \textbf{Cell Size.} The average size of the segmented cells found in the image.
\item \textbf{Zernike Shape.} A set of 30 features that describe the shape of cells using a basis of Zernike polynomials (order 0 to order 9).
\item \textbf{Expression Level.} A set of five features that measure the level of signal from the different cellular compartments in the image: DNA, mitochondria, endoplasmic reticulum, Golgi, cytoplasmic RNA, nucleoli, actin, and plasma membrane.
\end{itemize}
\vspace{-0.1cm}
We extract these features from a subset of images and compute the Spearman correlation between the features of real and generated images corresponding to the same molecule (see Supplementary Material for details). Due to space constraints, we show the mean of the correlation coefficients for the 30 Zernike shape features and the five expression level features.

\textbf{Other Evaluation Metrics.} In addition to these specialized metrics for biological images, we also evaluate our model using the following metrics that are more conventional for image generation tasks: 
\vspace{-0.5cm}
\begin{itemize}[leftmargin=*]
\itemsep0em 
    \item \textbf{Sliced Wasserstein Distance (SWD).} To assess the visual quality of the generated images, we consider the statistical similarity of image patches taken from multiple levels of a Laplacian pyramid representation of generated and real images, as described in \cite{karras2017progressive}.
    This metric compares the unconditional distributions of patches between generated and real images, but it does not take into account the correspondence between the generated image and the molecular information.
    \item \textbf{Correspondence Classification Accuracy (Corr).} To assess the correspondence between the generated images and the molecular information, we compute the accuracy of a pretrained correspondence classifier on the generated images. The classifier consists of a visual network and GNN that are trained on a binary classification task: detect whether the input cell image matches the input molecular intervention (positive sample) or whether they are mismatched (negative sample).
    The classifier 
    detects correctly matched pairs of images and molecules with an accuracy of $\sim$0.65 on real data (upper bound for our evaluation of generated images).
\end{itemize}

\begin{table}[t]
\centering
\small
\begin{tabular}{|c|c|c|c|c|c|c|c|}
\hline
Approach & Coverage & Cell Count & Cell Size & Zernike Shape & Exp. Level & SWD & Corr \\ \hline
CGAN \cite{gulrajani2017improved} & 7.0 & 4.8 & -2.9 & -3.9 & 7.4 & 5.65 & 56.6 \\ \hline
CGlow \cite{kingma2018glow} & -1.3 & 3.8 & 5.8 & 2.2 & 6.6 & 5.01 & 55.5 \\ \hline
w/o pyramid & 28.5 & 36.1 & 17.5 & 8.7 & 26.7 & 4.96 & 60.0 \\ \hline
w/o contrastive loss & 7.7 & 13.4 & 12.0 & 6.8 & 5.3 & \textbf{3.68} & 58.8 \\ \hline
Mol2Image (ours) & \textbf{44.6} & \textbf{54.4} & \textbf{27.5} & \textbf{15.8} & \textbf{37.3} & 4.63 & \textbf{63.2} \\ \hline
\end{tabular}
\vspace{0.1cm}
\caption{Evaluation of Mol2Image (our model) vs. the baselines on images generated from molecules from the training set. ``Coverage", ``Cell Count", ``Cell Size", ``Zernike Shape", ``Exp. Level" measure Spearman correlation coefficients ($\times 10^2$) between features from a subset of real and generated images; higher is better. ``Corr" represents correspondence classification accuracy of a pretrained model; higher is better and ground truth (upper bound) achieves $65.0$. ``SWD" is the sliced Wasserstein distance metric ($\times 10^{-2}$) from \cite{karras2017progressive}; lower is better. See text for details.}
\label{table:image-metrics}
\end{table}

\begin{table}[]
\centering
\small
\begin{tabular}{|c|c|c|c|c|c|c|c|}
\hline
Approach & Coverage & Cell Count & Cell Size & Zernike Shape & Exp. Level & SWD & Corr \\ \hline
CGAN \cite{gulrajani2017improved} & 6.4 & 1.9 & -1.5 & -1.0 & 9.2 & 5.60 & 56.1 \\ \hline
CGlow \cite{kingma2018glow} & 3.1 & -3.7 & -3.0 & -3.1 & 3.7 & 5.40 & 54.5 \\ \hline
w/o pyramid & 9.2 & 1.7 & \textbf{12.9} & \textbf{6.1} & 8.6 & 4.20 & 59.1 \\ \hline
w/o contrastive loss & 5.0 & 9.1 & 6.1 & 2.9 & 9.2 & \textbf{3.41} & 55.7 \\ \hline
Mol2Image (ours) & \textbf{15.8} & \textbf{19.7} & 11.0 & 4.9 & \textbf{13.4} & 4.27 & \textbf{62.6} \\ \hline
\end{tabular}
\vspace{0.1cm}
\caption{Same as Table \ref{table:image-metrics}, but evaluated on images generated from held-out molecules. Ground truth (upper bound) achieves $64.2$ on the correspondence classification accuracy (Corr) metric.}
\label{table:image-metrics-held-out}
\end{table}

\textbf{Baselines and Ablations.} Since molecule-to-image synthesis is a novel task, we develop our own baselines based on well-established generative models and perform ablations to determine the benefit of our approach. Since not all of the methods are capable of generating high-quality images at full $512\times 512$ resolution, we compare all of the model results at $64\times 64$ spatial resolution.
\begin{itemize}[leftmargin=*]
\itemsep0em 
    \item \textbf{Baseline: Conditional GAN with Graph Neural Network (CGAN).} We train a CGAN such that a generator network $G$ is trained to generate images conditioned on the corresponding molecule, $y$. Both the generator $G$ and discriminator $D$ are conditioned on the molecular representation $g(y)$ learned by the same GNN as above. We use a Wasserstein GAN trained with a gradient penalty \cite{gulrajani2017improved}. This variant is able to consistently produce qualitatively realistic images in the unconditional setting, in agreement with previous generative models for cell image data \cite{osokin2017gans}. 
    \item \textbf{Baseline: Conditional Glow with Graph Neural Network (CGlow).} Since our model is an improved flow-based model for conditional generation, we develop a baseline approach based on existing work that is a straightforward extension of Glow to the conditional setting \cite{kingma2018glow}. Specifically, this baseline model conditions the distribution of latent variables introduced at every level of the multi-scale architecture on the output of the graph neural network and optimizes the conditional log-likelihood with respect to the model parameters. Alternatively, this model can be seen as an ablation of our model without pyramid architecture or contrastive training.
    \item \textbf{Ablation: Mol2Image without Pyramid Architecture (w/o pyramid).} We train our model without the framework of the image pyramid for separately generating features at different scales. Instead, we directly generate the full resolution image.
    \item \textbf{Ablation: Mol2Image without Contrastive Learning (w/o contrastive loss).} We train our model without using contrastive loss to maximize the mutual information between the latent variables of the image and the embeddings extracted by the graph neural network.
\end{itemize}


\textbf{Results.} Tables \ref{table:image-metrics} and \ref{table:image-metrics-held-out} show the results of our model in comparison to the baselines. Our conditional flow-based generative model, which is trained with the proposed pyramid architecture and the contrastive loss, outperforms the baselines in generating cell images that reflect the effects of the molecular interventions. Table \ref{table:image-metrics} shows that our model performs well on generating cell images conditioned on molecules that were observed during training. Table \ref{table:image-metrics-held-out} shows that our model generalizes better than the baselines to molecules that were held-out from the training set. 

\textbf{Effect of Contrastive Loss.}
Our training strategy, which uses contrastive loss to maximize the mutual information between the image latent variables and the molecular embedding, is essential for effective generation of images conditioned on the molecular intervention. In particular, there is much lower correspondence between the images and the molecular intervention when contrastive learning is omitted. This result holds both in the case that we use the image pyramid framework (i.e., compare "Mol2Image" with "w/o contrastive learning") and in the case that we directly generate 64 x 64 images using the standard multi-scale architecture (i.e., compare "w/o pyramid architecture" with "CGlow"). 
This demonstrates that contrastive learning can provide a strong signal for learning the relation between the image and the conditioning information for generative modeling, in the absence of categorical labels that can be used in a supervised framework. 
On the other hand, contrastive loss does not appear to improve the unconditional quality of generated images (based on SWD).

\textbf{Effect of Pyramid Framework.}
We proposed the pyramid structure to generate image features at different spatial resolutions, which is important to disentangle higher level features (e.g., cell distribution) from lower level features (e.g., cell shape), and to allow our model to scale to high-resolution cell images (512 x 512). Interestingly, we find that the image pyramid framework also improves the conditional generation of 64 x 64 images compared to the baseline model that directly generates images of this size (i.e., compare "Mol2Image" to "w/o pyramid"). We hypothesize that this is because it is more efficient and easier to learn the relation between images and conditions when starting with the low-resolution images at the bottom of the image pyramid. Consistent with our observations, previous works have reported that training GANs starting from lower-resolution images \cite{karras2017progressive} or using an image pyramid \cite{denton2015deep} is more effective than training directly on full-resolution images.

\textbf{Qualitative Examples.}
Figure \ref{fig:generated-examples} shows a qualitative comparison between the baselines (CGAN, CGlow) and our method on generating images conditioned on molecular structure. The generated images from our method (Figure \ref{fig:generated-examples}, row 3) more closely reflect the real effect of the intervention (Figure \ref{fig:generated-examples}, row 4) compared to other methods, both in terms of cell morphology and in terms of channel intensities (representing expression of different cellular components). More qualitative examples (including full-resolution $512 \times 512$ images) are provided in the Supplementary Material.

\begin{table}[]
\small
\centering
\begin{tabular}{|c|c|c|c|c|}
\hline
Molecular Embedding & Random & Morgan Fingerprint & w/o Contrastive Loss & Mol2Image (ours) \\ \hline
Mean AUC & 0.569 & 0.645 & 0.675 & \textbf{0.810} \\ \hline
Mean AUC (Held-Out) & 0.578 & 0.665 & 0.675 & \textbf{0.683} \\ \hline
\end{tabular}
\vspace{0.1cm}
\caption{Evaluation of molecular embeddings on predicting morphological labels. Higher AUC is better. ``Random" refers to embeddings from a randomly initialized GNN. ``Held-out" refers to held-out molecules from the training set. For reference, a fully-supervised model (in which the parameters of the graph neural network are trained) achieves an AUC of 0.702 on held-out molecules.}
\label{table:mol-embeddings}
\end{table}

\textbf{Analysis of Molecular Embeddings.}
Since our method performs well at generating cell images conditioned on molecular interventions, we hypothesize that the GNN learns a molecular representation that reflects the morphological features of the cell image. To determine whether the molecular embeddings are linearly separable based on the morphology they induce in treated cells, we train a linear classifier to predict a subset of 14 features curated from the morphological analysis of Bray \etal \cite{bray2017dataset} (see the Supplementary Material). For comparison, we consider embeddings from a randomly initialized GNN, Morgan/circular fingerprints \cite{rogers2010extended}, and an ablation of our model trained without contrastive loss. Table \ref{table:mol-embeddings} shows the average AUC of the various embeddings on this task. The results suggest that our method learns molecular embeddings that are linearly-separable based on morphological properties of the treated cells (Table \ref{table:mol-embeddings}, Row 1), and that the learned embeddings can also generalize to previously unseen molecules (Table \ref{table:mol-embeddings}, Row 2).

\section{Discussion}

We have developed a new multi-scale flow-based architecture and training strategy for molecule-to-image synthesis and demonstrated the benefits of our approach on new evaluation metrics tailored to biological cell image generation. Our work represents a first step towards image-based virtual screening of chemicals and lays the groundwork for studying the shared information in molecular structures and perturbed cell morphology. 
A promising avenue for future work is integrating side information (e.g., known chemical properties, drug dosage) to impose constraints on the molecular embedding space and improve generalization to previously unseen molecules. Furthermore, even though we have focused on molecule-to-image synthesis in this paper, our contributions to flow-based models can potentially be applied in other contexts, e.g., text-to-image synthesis \cite{reed2016generative}.


\section*{Acknowledgements}
Karren Dai Yang was supported by an NSF Graduate Research Fellowship and ONR (N00014-18-1-2765). Alex X. Lu was funded by a pre-doctoral award from the National Science and Engineering Research Council. Regina Barzilay and Tommi Jaakkola were partially supported by the MLPDS Consortium and the DARPA AMD program. Caroline Uhler was partially supported by NSF (DMS-1651995), ONR (N00014-17-1-2147 and N00014-18-1-2765), IBM, and a Simons Investigator Award.

{\small
\bibliographystyle{ieee_fullname}
\bibliography{egbib}

\begin{thebibliography}{10}\itemsep=-1pt

\bibitem{ardizzone2019guided}
Lynton Ardizzone, Carsten L{\"u}th, Jakob Kruse, Carsten Rother, and Ullrich
  K{\"o}the.
\newblock Guided image generation with conditional invertible neural networks.
\newblock {\em arXiv preprint arXiv:1907.02392}, 2019.

\bibitem{bray2017dataset}
Mark-Anthony Bray, Sigrun~M Gustafsdottir, Mohammad~H Rohban, Shantanu Singh,
  Vebjorn Ljosa, Katherine~L Sokolnicki, Joshua~A Bittker, Nicole~E Bodycombe,
  Vlado Dan{\v{c}}{\'\i}k, Thomas~P Hasaka, et~al.
\newblock A dataset of images and morphological profiles of 30 000
  small-molecule treatments using the cell painting assay.
\newblock {\em Gigascience}, 6(12):giw014, 2017.

\bibitem{bray2016cell}
Mark-Anthony Bray, Shantanu Singh, Han Han, Chadwick~T Davis, Blake Borgeson,
  Cathy Hartland, Maria Kost-Alimova, Sigrun~M Gustafsdottir, Christopher~C
  Gibson, and Anne~E Carpenter.
\newblock Cell painting, a high-content image-based assay for morphological
  profiling using multiplexed fluorescent dyes.
\newblock {\em Nature protocols}, 11(9):1757, 2016.

\bibitem{breinig2015chemical}
Marco Breinig, Felix~A Klein, Wolfgang Huber, and Michael Boutros.
\newblock A chemical--genetic interaction map of small molecules using
  high-throughput imaging in cancer cells.
\newblock {\em Molecular systems biology}, 11(12), 2015.

\bibitem{caicedo2018weakly}
Juan~C Caicedo, Claire McQuin, Allen Goodman, Shantanu Singh, and Anne~E
  Carpenter.
\newblock Weakly supervised learning of single-cell feature embeddings.
\newblock In {\em Proceedings of IEEE Conference on Computer Vision and Pattern
  Recognition (CVPR)}, pages 9309--9318, 2018.

\bibitem{caicedo2016applications}
Juan~C Caicedo, Shantanu Singh, and Anne~E Carpenter.
\newblock Applications in image-based profiling of perturbations.
\newblock {\em Current opinion in biotechnology}, 39:134--142, 2016.

\bibitem{dai2016discriminative}
Hanjun Dai, Bo Dai, and Le Song.
\newblock Discriminative embeddings of latent variable models for structured
  data.
\newblock In {\em International Conference on Machine Learning}, pages
  2702--2711, 2016.

\bibitem{denton2015deep}
Emily~L Denton, Soumith Chintala, Rob Fergus, et~al.
\newblock Deep generative image models using a￼ laplacian pyramid of
  adversarial networks.
\newblock In {\em Advances in neural information processing systems}, pages
  1486--1494, 2015.

\bibitem{dinh2014nice}
Laurent Dinh, David Krueger, and Yoshua Bengio.
\newblock Nice: Non-linear independent components estimation.
\newblock {\em arXiv preprint arXiv:1410.8516}, 2014.

\bibitem{dinh2016density}
Laurent Dinh, Jascha Sohl-Dickstein, and Samy Bengio.
\newblock Density estimation using real nvp.
\newblock {\em arXiv preprint arXiv:1605.08803}, 2016.

\bibitem{duvenaud2015convolutional}
David~K Duvenaud, Dougal Maclaurin, Jorge Iparraguirre, Rafael Bombarell,
  Timothy Hirzel, Al{\'a}n Aspuru-Guzik, and Ryan~P Adams.
\newblock Convolutional networks on graphs for learning molecular fingerprints.
\newblock In {\em Advances in neural information processing systems}, pages
  2224--2232, 2015.

\bibitem{eggert2013and}
Ulrike~S Eggert.
\newblock The why and how of phenotypic small-molecule screens.
\newblock {\em Nature chemical biology}, 9(4):206, 2013.

\bibitem{gilmer2017neural}
Justin Gilmer, Samuel~S Schoenholz, Patrick~F Riley, Oriol Vinyals, and
  George~E Dahl.
\newblock Neural message passing for quantum chemistry.
\newblock {\em arXiv preprint arXiv:1704.01212}, 2017.

\bibitem{goldsborough2017cytogan}
Peter Goldsborough, Nick Pawlowski, Juan~C Caicedo, Shantanu Singh, and Anne
  Carpenter.
\newblock Cytogan: generative modeling of cell images.
\newblock {\em bioRxiv}, page 227645, 2017.

\bibitem{goodfellow2014generative}
Ian Goodfellow, Jean Pouget-Abadie, Mehdi Mirza, Bing Xu, David Warde-Farley,
  Sherjil Ozair, Aaron Courville, and Yoshua Bengio.
\newblock Generative adversarial nets.
\newblock In {\em Advances in neural information processing systems}, pages
  2672--2680, 2014.

\bibitem{gori2005graph}
Marco Gori, Gabriele Monfardini, and Franco Scarselli.
\newblock A new model for learning in graph domains.
\newblock In {\em Neural Networks, 2005. IJCNN'05. Proceedings. 2005 IEEE
  International Joint Conference on}, volume~2, pages 729--734. IEEE, 2005.

\bibitem{gulrajani2017improved}
Ishaan Gulrajani, Faruk Ahmed, Martin Arjovsky, Vincent Dumoulin, and Aaron~C
  Courville.
\newblock Improved training of wasserstein gans.
\newblock In {\em Advances in Neural Information Processing Systems (NeurIPS)},
  pages 5767--5777, 2017.

\bibitem{hamilton2017inductive}
William~L Hamilton, Rex Ying, and Jure Leskovec.
\newblock Inductive representation learning on large graphs.
\newblock {\em arXiv preprint arXiv:1706.02216}, 2017.

\bibitem{ho2019flow++}
Jonathan Ho, Xi Chen, Aravind Srinivas, Yan Duan, and Pieter Abbeel.
\newblock Flow++: Improving flow-based generative models with variational
  dequantization and architecture design.
\newblock {\em arXiv preprint arXiv:1902.00275}, 2019.

\bibitem{hofmarcher2019accurate}
Markus Hofmarcher, Elisabeth Rumetshofer, Djork-Arne Clevert, Sepp Hochreiter,
  and Günter Klambauer.
\newblock Accurate prediction of biological assays with high-throughput
  microscopy images and convolutional networks.
\newblock {\em Journal of chemical information and modeling}, 59(3):1163--1171,
  2019.

\bibitem{hou2017unsupervised}
Le Hou, Ayush Agarwal, Dimitris Samaras, Tahsin~M Kurc, Rajarsi~R Gupta, and
  Joel~H Saltz.
\newblock Unsupervised histopathology image synthesis.
\newblock {\em arXiv preprint arXiv:1712.05021}, 2017.

\bibitem{karras2017progressive}
Tero Karras, Timo Aila, Samuli Laine, and Jaakko Lehtinen.
\newblock Progressive growing of gans for improved quality, stability, and
  variation.
\newblock {\em ICLR}, 2018.

\bibitem{kearnes2016molecular}
Steven Kearnes, Kevin McCloskey, Marc Berndl, Vijay Pande, and Patrick Riley.
\newblock Molecular graph convolutions: moving beyond fingerprints.
\newblock {\em Journal of computer-aided molecular design}, 30(8):595--608,
  2016.

\bibitem{kingma2014adam}
Diederik~P Kingma and Jimmy Ba.
\newblock Adam: A method for stochastic optimization.
\newblock {\em Proceedings of the International Conference on Learning
  Representations (ICLR)}, 2014.

\bibitem{kingma2018glow}
Durk~P Kingma and Prafulla Dhariwal.
\newblock Glow: Generative flow with invertible 1x1 convolutions.
\newblock In {\em Advances in Neural Information Processing Systems}, pages
  10215--10224, 2018.

\bibitem{kingma2013auto}
Diederik~P Kingma and Max Welling.
\newblock Auto-encoding variational bayes.
\newblock {\em arXiv preprint arXiv:1312.6114}, 2013.

\bibitem{kipf2016semi}
Thomas~N Kipf and Max Welling.
\newblock Semi-supervised classification with graph convolutional networks.
\newblock {\em International Conference on Learning Representations}, 2017.

\bibitem{kondo2019flow}
Ruho Kondo, Keisuke Kawano, Satoshi Koide, and Takuro Kutsuna.
\newblock Flow-based image-to-image translation with feature disentanglement.
\newblock In {\em Advances in Neural Information Processing Systems}, pages
  4170--4180, 2019.

\bibitem{lei2017deriving}
Tao Lei, Wengong Jin, Regina Barzilay, and Tommi Jaakkola.
\newblock Deriving neural architectures from sequence and graph kernels.
\newblock {\em International Conference on Machine Learning}, 2017.

\bibitem{li2015gated}
Yujia Li, Daniel Tarlow, Marc Brockschmidt, and Richard Zemel.
\newblock Gated graph sequence neural networks.
\newblock {\em arXiv preprint arXiv:1511.05493}, 2015.

\bibitem{liu2019conditional}
Rui Liu, Yu Liu, Xinyu Gong, Xiaogang Wang, and Hongsheng Li.
\newblock Conditional adversarial generative flow for controllable image
  synthesis.
\newblock In {\em Proceedings of the IEEE Conference on Computer Vision and
  Pattern Recognition}, pages 7992--8001, 2019.

\bibitem{ljosa2013comparison}
Vebjorn Ljosa, Peter~D Caie, Rob Ter~Horst, Katherine~L Sokolnicki, Emma~L
  Jenkins, Sandeep Daya, Mark~E Roberts, Thouis~R Jones, Shantanu Singh,
  Auguste Genovesio, et~al.
\newblock Comparison of methods for image-based profiling of cellular
  morphological responses to small-molecule treatment.
\newblock {\em Journal of biomolecular screening}, 18(10):1321--1329, 2013.

\bibitem{loo2009approach}
Lit-Hsin Loo, Hai-Jui Lin, Robert~J Steininger~III, Yanqin Wang, Lani~F Wu, and
  Steven~J Altschuler.
\newblock An approach for extensibly profiling the molecular states of cellular
  subpopulations.
\newblock {\em Nature methods}, 6(10):759, 2009.

\bibitem{louizos2017multiplicative}
Christos Louizos and Max Welling.
\newblock Multiplicative normalizing flows for variational bayesian neural
  networks.
\newblock In {\em Proceedings of the 34th International Conference on Machine
  Learning-Volume 70}, pages 2218--2227. JMLR. org, 2017.

\bibitem{lu2019cells}
Alex Lu, Amy Lu, Wiebke Schormann, Marzyeh Ghassemi, David Andrews, and Alan
  Moses.
\newblock The cells out of sample (coos) dataset and benchmarks for measuring
  out-of-sample generalization of image classifiers.
\newblock In {\em Advances in Neural Information Processing Systems}, pages
  1852--1860, 2019.

\bibitem{lu2019structured}
You Lu and Bert Huang.
\newblock Structured output learning with conditional generative flows.
\newblock {\em arXiv preprint arXiv:1905.13288}, 2019.

\bibitem{mahmood2018unsupervised}
Faisal Mahmood, Richard Chen, and Nicholas~J Durr.
\newblock Unsupervised reverse domain adaptation for synthetic medical images
  via adversarial training.
\newblock {\em IEEE transactions on medical imaging}, 37(12):2572--2581, 2018.

\bibitem{mattiazzi2020systematic}
Mojca Mattiazzi~Usaj, Nil Sahin, Helena Friesen, Carles Pons, Matej Usaj, Myra
  Paz~D Masinas, Ermira Shuteriqi, Aleksei Shkurin, Patrick Aloy, Quaid Morris,
  et~al.
\newblock Systematic genetics and single-cell imaging reveal widespread
  morphological pleiotropy and cell-to-cell variability.
\newblock {\em Molecular systems biology}, 16(2):e9243, 2020.

\bibitem{mcquin2018cellprofiler}
Claire McQuin, Allen Goodman, Vasiliy Chernyshev, Lee Kamentsky, Beth~A Cimini,
  Kyle~W Karhohs, Minh Doan, Liya Ding, Susanne~M Rafelski, Derek Thirstrup,
  et~al.
\newblock Cellprofiler 3.0: Next-generation image processing for biology.
\newblock {\em PLoS biology}, 16(7), 2018.

\bibitem{niepert2016learning}
Mathias Niepert, Mohamed Ahmed, and Konstantin Kutzkov.
\newblock Learning convolutional neural networks for graphs.
\newblock In {\em International Conference on Machine Learning}, pages
  2014--2023, 2016.

\bibitem{okagaki2010cryptococcal}
Laura~H Okagaki, Anna~K Strain, Judith~N Nielsen, Caroline Charlier, Nicholas~J
  Baltes, Fabrice Chr{\'e}tien, Joseph Heitman, Fran{\c{c}}oise Dromer, and
  Kirsten Nielsen.
\newblock Cryptococcal cell morphology affects host cell interactions and
  pathogenicity.
\newblock {\em PLoS pathogens}, 6(6), 2010.

\bibitem{oord2018representation}
Aaron van~den Oord, Yazhe Li, and Oriol Vinyals.
\newblock Representation learning with contrastive predictive coding.
\newblock {\em arXiv preprint arXiv:1807.03748}, 2018.

\bibitem{osokin2017gans}
Anton Osokin, Anatole Chessel, Rafael~E Carazo~Salas, and Federico Vaggi.
\newblock Gans for biological image synthesis.
\newblock In {\em Proceedings of IEEE International Conference on Computer
  Vision (ICCV)}, pages 2233--2242, 2017.

\bibitem{perlman2004multidimensional}
Zachary~E Perlman, Michael~D Slack, Yan Feng, Timothy~J Mitchison, Lani~F Wu,
  and Steven~J Altschuler.
\newblock Multidimensional drug profiling by automated microscopy.
\newblock {\em Science}, 306(5699):1194--1198, 2004.

\bibitem{reddy2007virtual}
A~Srinivas Reddy, S~Priyadarshini Pati, P~Praveen Kumar, HN Pradeep, and
  G~Narahari Sastry.
\newblock Virtual screening in drug discovery-a computational perspective.
\newblock {\em Current Protein and Peptide Science}, 8(4):329--351, 2007.

\bibitem{reed2016generative}
Scott Reed, Zeynep Akata, Xinchen Yan, Lajanugen Logeswaran, Bernt Schiele, and
  Honglak Lee.
\newblock Generative adversarial text to image synthesis.
\newblock {\em arXiv preprint arXiv:1605.05396}, 2016.

\bibitem{rezende2015variational}
Danilo~Jimenez Rezende and Shakir Mohamed.
\newblock Variational inference with normalizing flows.
\newblock {\em arXiv preprint arXiv:1505.05770}, 2015.

\bibitem{rogers2010extended}
David Rogers and Mathew Hahn.
\newblock Extended-connectivity fingerprints.
\newblock {\em Journal of chemical information and modeling}, 50(5):742--754,
  2010.

\bibitem{rohban2017systematic}
Mohammad~Hossein Rohban, Shantanu Singh, Xiaoyun Wu, Julia~B Berthet,
  Mark-Anthony Bray, Yashaswi Shrestha, Xaralabos Varelas, Jesse~S Boehm, and
  Anne~E Carpenter.
\newblock Systematic morphological profiling of human gene and allele function
  via cell painting.
\newblock {\em Elife}, 6:e24060, 2017.

\bibitem{scarselli2009graph}
Franco Scarselli, Marco Gori, Ah~Chung Tsoi, Markus Hagenbuchner, and Gabriele
  Monfardini.
\newblock The graph neural network model.
\newblock {\em IEEE Transactions on Neural Networks}, 20(1):61--80, 2009.

\bibitem{shamir2011assessing}
L Shamir.
\newblock Assessing the efficacy of low-level image content descriptors for
  computer-based fluorescence microscopy image analysis.
\newblock {\em Journal of microscopy}, 243(3):284--292, 2011.

\bibitem{shoichet2004virtual}
Brian~K Shoichet.
\newblock Virtual screening of chemical libraries.
\newblock {\em Nature}, 432(7019):862--865, 2004.

\bibitem{sun2019dual}
Haoliang Sun, Ronak Mehta, Hao~H Zhou, Zhichun Huang, Sterling~C Johnson, Vivek
  Prabhakaran, and Vikas Singh.
\newblock Dual-glow: Conditional flow-based generative model for modality
  transfer.
\newblock In {\em Proceedings of the IEEE International Conference on Computer
  Vision}, pages 10611--10620, 2019.

\bibitem{swinney2011were}
David~C Swinney and Jason Anthony.
\newblock How were new medicines discovered?
\newblock {\em Nature reviews Drug discovery}, 10(7):507--519, 2011.

\bibitem{velickovic2017graph}
Petar Velickovic, Guillem Cucurull, Arantxa Casanova, Adriana Romero, Pietro
  Lio, and Yoshua Bengio.
\newblock Graph attention networks.
\newblock {\em arXiv preprint arXiv:1710.10903}, 2017.

\bibitem{walters1998virtual}
W~Patrick Walters, Matthew~T Stahl, and Mark~A Murcko.
\newblock Virtual screening—an overview.
\newblock {\em Drug discovery today}, 3(4):160--178, 1998.

\bibitem{winkler2019learning}
Christina Winkler, Daniel Worrall, Emiel Hoogeboom, and Max Welling.
\newblock Learning likelihoods with conditional normalizing flows.
\newblock {\em arXiv preprint arXiv:1912.00042}, 2019.

\bibitem{xu2018powerful}
Keyulu Xu, Weihua Hu, Jure Leskovec, and Stefanie Jegelka.
\newblock How powerful are graph neural networks?
\newblock {\em arXiv preprint arXiv:1810.00826}, 2018.

\bibitem{yang2019analyzing}
Kevin Yang, Kyle Swanson, Wengong Jin, Connor Coley, Philipp Eiden, Hua Gao,
  Angel Guzman-Perez, Timothy Hopper, Brian Kelley, Miriam Mathea, et~al.
\newblock Analyzing learned molecular representations for property prediction.
\newblock {\em Journal of chemical information and modeling}, 59(8):3370--3388,
  2019.

\bibitem{yi2019generative}
Xin Yi, Ekta Walia, and Paul Babyn.
\newblock Generative adversarial network in medical imaging: A review.
\newblock {\em Medical image analysis}, page 101552, 2019.

\end{thebibliography}
}

\appendix


\section*{Supplementary material}

# Appendix

## A  Analysis of Molecular Embeddings

As described in the main text, to evaluate the molecular embedding space learned by our graph neural network, we train a linear classifier to predict a subset of morphological features. We curate the labels for this task as follows. The dataset of Bray *et al.* [1] provides measured values of the following morphological features for every cell image in their dataset: area, compactness, eccentricity, form factor, major axis length, minor axis length, radius, perimeter, solidity, and cell count. To each small molecule, we assign a continuous-valued vector representing the mean values of these features observed in cells treated with that molecule. Direct prediction of these values is not a meaningful task because the amount of intra-molecular variability is high relative to the inter-molecular variability; much of the variability in the features may be naturally occurring due to stochasticity in cell growth and is not explained by the molecular perturbation. Therefore, we predict instead the presence of atypical morphology caused by a molecule. We convert these continuous values to binary labels – 1 if the value is in the top or bottom 1% / 5% / 10% of the values for its class, and 0 otherwise – and train a logistic regression model to perform multi-task binary classification. The results in Table 3 of the main text show that the molecular embeddings learned by our graph neural network reflect morphological properties of treated cells and enable linear separation of molecules that cause atypical morphological features. Upon acceptance of the work, we will release the molecular metadata and splits used in our morphology prediction task.

## B  CellProfiler Evaluation

CellProfiler [2] is a standard open-source software used for segmenting cells/nuclei and quantifying specific morphological features. The segmentation of nuclei and cells occurs in two steps: (1) thresholding is performed to identify the nuclei from the DNA stain, and (2) the nuclei are used as reference points for determining boundaries between cells and identifying cell objects. Once the cells are identified, multiple pipelines are available to measure shape and intensity features within each cell. To evaluate the generated images from our model, we extract morphological features for a subset of generated and held-out images and compute the correlation coefficient between the features of generated and real images. To increase the range of phenotypes within the evaluated subset, we focus our evaluation on molecules that are more likely to cause a morphological change in cells, based on the morphology criterion used in Section A. Upon acceptance of the work, we will release the molecular metadata and CellProfiler pipeline used for evaluation.

## C  Additional Qualitative Examples

See Supplemental Figures 1 and 2 for examples of full-resolution cell images generated by our method.

## References

[1] Mark-Anthony Bray, Sigrun M Gustafsdottir, Mohammad H Rohban, Shantanu Singh, Vebjorn Ljosa, Katherine L Sokolnicki, Joshua A Bittker, Nicole E Bodycombe, Vlado Dančík, Thomas P Hasaka, et al. A dataset of images and morphological profiles of 30 000 small-molecule treatments using the cell painting assay. *Gigascience*, 6(12):giw014, 2017.

[2] Claire McQuin, Allen Goodman, Vasiliy Chernyshev, Lee Kamentsky, Beth A Cimini, Kyle W Karhohs, Minh Doan, Liya Ding, Susanne M Rafelski, Derek Thirstrup, et al. Cellprofiler 3.0: Next-generation image processing for biology. *PLoS biology*, 16(7), 2018.

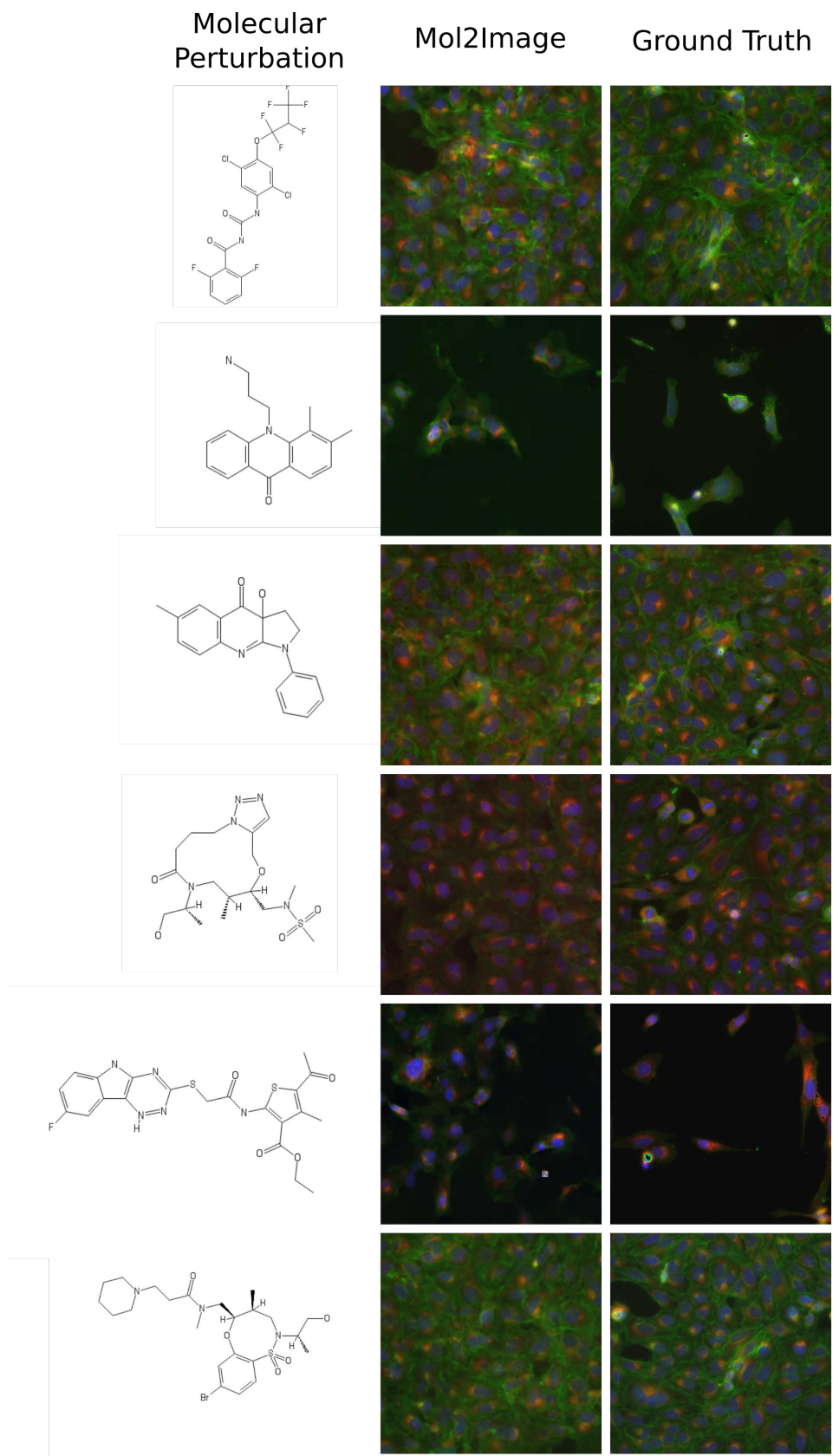

Figure 1: Examples of full-resolution cell images generated by our method.

| Molecular Perturbation | Mol2Image | Ground Truth |

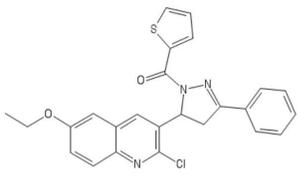

Figure 2: Examples of full-resolution cell images generated by our method.



\end{document}